\documentclass[a4paper]{jpconf}
\usepackage{graphicx}
\usepackage{amsmath}  % math typesetting
\pdfoutput=1
\usepackage{amssymb}    % for math symbol 
\usepackage{graphicx}  % plots
\usepackage{hyperref}
\hypersetup{colorlinks=true,linkcolor=blue,urlcolor=blue}
\begin{document}
\title{Drag Control by Hydrogen Injection in Shocked Stagnation Zone of Blunt Nose}

\author{Ashish Vashishtha$^{\boldsymbol{\mathsf{1}}}$,
Dean Callaghan$^{\boldsymbol{\mathsf{1}}}$ and
Cathal Nolan$^{\boldsymbol{\mathsf{1}}}$}

\address{$^{\mathsf{1}}$Department of Aerospace, Mechanical and Electronic Engineering, Institute of Technology Carlow, IRELAND.}

\ead{ashish.vashishtha@itcarlow.ie}

\begin{abstract}
The main motivation of the current study is to propose a high-pressure hydrogen injection as an hybrid active flow control technique in order to manipulate the flow-field in front of a blunt nose during hypersonic flight. Hydrogen injection can lead to self-ignition under the right environment conditions in a stagnation zone, and may cause thermal heat addition through combustion and provide the counterjet effect together by pushing bow shock upstream. The axisymmetric numerical simulations for the hemispherical blunt nose are performed at a Mach 6 freestream flow with 10000 Pa pressure and 293 K temperature. The sonic and supersonic hydrogen and air injections are compared for drag reduction at the same stagnation pressure ratio $PR$ and momentum ratio ($R_{MA}$). The sonic air and hydrogen injection scenarios show similar performance in terms of drag reduction and similar SPM flow features, but hydrogen injection has a mass flow rate 3.76 times lower than air. Supersonic hydrogen injection at $M_j$ 2.94 behaves differently than supersonic air injection and can achieve up to 60 \% drag reduction at lower PR and LPM mode with lower mass flow rate. Additionally, air injection achieves a drag reduction of 40 \% in SPM mode at higher PR with very high mass flow rate.
\end{abstract}
\section{Introduction}
The high-speed flight of an air-vehicle in supersonic and hypersonic flow regimes have the formation of shockwaves in front of the nose. The high pressure stagnation zone in the frontal shock wave can cause high wave drag, which leads to higher total drag of the vehicle. In high supersonic and hypersonic flow regimes, a major amount of flow kinetic energy will be converted to internal energy of the gas across the shock. It causes high aerodynamic heating in highly compressed stagnation zone\cite{erath}. Many active and passive drag control methods have been extensively studied by manipulation of the stagnation zone in supersonic and hypersonic flows such as mechanical spike, counterjet flow and energy deposition \cite{reviewp}. Among them, mechanical spike is a passive drag control method and counterjet flow and energy deposition are active drag control methods. Breathing blunt nose is another passive drag control methods studied in supersonic \cite{vashish} and hypersonic \cite{watanabe} flow regimes. The majority of the drag control methods work by manipulation of the frontal shock wave either by pushing away from the body as in the case of spike, counterjet flow and energy deposition or bringing closer the frontal shock and manipulating the base drag in breathing blunt nose. Among the passive flow controls, the main disadvantages of mechanical spike are unsteady flow-fields for certain length of spikes and it adds additional length to the vehicle which may produce additional side forces. The breathing blunt nose method was found to be promising but it may be required to further study the effectiveness in high enthalpy flows and heating control in internal flow. Among active control methods, the energy deposition method may require a complicated energy source device onboard either for electric discharge, laser or microwave based energy deposition. The counterjet flow have been found one of the effective technique to control the drag, which can be used to reduce the drag, when small mass flow is injected in counterjet flow manner or to increase drag (increase thrust, while re-entry as in retro-propulsion), when higher mass flow (mostly exhaust from engine) is ejected. However, most studies are performed with either air or non-reactive gases in drag reduction counterjet flow controls. Warren \cite{warren} studied helium and nitrogen injection and found that helium (lighter) gas was superior to reduce peak at pressure reattachment. A number of studies investigated hybrid drag control methods for example lateral jet injection from spike \cite{jiang} and electrical discharge from the spike tip \cite{kuo}. However, to author's knowledge, there are no hybrid studies involving energy addition (in the stagnation zone) and counterjet flow in order to achieve drag control. This study aims to investigate the feasibility of using hydrogen as counterjet flow gas in order to control the drag and compare the drag reduction with respect to counterjet air operating under the same conditions. The low ignition energy required to ignite hydrogen, its high diffusivity and being a lightest gas are the main reasons behind using hydrogen as a counterjet flow jet to control drag. 
%Further, it is also envisioned that hydrogen ignition in stagnation zone will start as deflagration flame because of self-ignition and depends on appropriate mixing, pressure and temperature, it may also be possible that hydrogen deflagration flame may turn into local detonation or high speed deflagration flame, which can push the frontal bow shock further away from the body. In this way hydrogen injection in the stagnation zone may work as a hybrid device of fluidic spike as well as thermal energy addition to stagnation zone. However, there may be concerns of high heating due to the high temperature flame in front of blunt nose. It is thought however, that a stagnation flame may form between the shock wave and surface of the blunt nose, which may not increase heating loads in hypersonic flow.
However, there may be concerns of high heating due to the high temperature flame in front of blunt nose. The current study is focused only on drag control by hydrogen injection in the stagnation zone. A complimentary heat transfer study will be carried out in future work. %Other system level advantages of drag control by hydrogen injection in the stagnation zone are: 1) hydrogen can be stored in lerger quantities as compared to air of same weight, 2) there is no additional source required in order to ignite, injected hydrogen can mix with air in the stagnation zone and utilize the high pressure and temperature zone in stagnation zone to self-ignite, 3) helium can be another alternative (lighter and inert gas) to hydrogen, but it may require additional storage for helium, however hydrogen may be on-board for future hypersonic transport applications. There may also be concerns about diffusion of unburned hydrogen through the shock layer, which can be further optimized with pulsed injection of hydrogen for higher penetration.

To the best of author's knowledge, the current study is the first that targets drag control by direct hydrogen injection as counterjet flow in the stagnation zone. Srinivasan et al. \cite{srinivasan} have studied the hydrogen injection in a spiked tipped body and found 35-50 \% drag reduction. Zhao et al.\cite{zhao} have recently simulated non-equilibrium flow at hypersonic and shock induced combustion in hydrogen air mixture for spherical blunt nose as well as hydrogen injection (found flow behaviour close to air injection) from a numerical tool's capability point of view. Many studies have been performed by researchers for oblique detonation wave propagation in a hydrogen-air freestream premixed mixture \cite{qin} and traverse hot-jet induced detonation propagation\cite{cai} for a pressure gain propulsion system. It is well established that the counterjet flow may behave in short (SPM) and long penetration modes (LPM) \cite{hayashi}, which depends on stagnation pressure ratio of counterjet flow to freestream flow and nozzle design. Sonic (convergent) counterjet nozzles usually exhibit short penetration modes at most of operating pressures, while supersonic (convergent-divergent) counterjet nozzles exhibit long penetration mode (LPM) or short penetration modes (SPM), depends on operating stagnation pressure. In LPM mode operation, the shock will be moved at longer upstream distances, which can lead to higher drag reduction. In short penetration mode, the Mach disk can be formed near central region and a low pressure region can be formed near the blunt body, which can cause low drag on the body. The current study is based on axisymmeytric two-dimensional simulation of a counterjet flow injection from a hemispherical blunt nose using the open-source platform OpenFOAM \cite{weller} at hypersonic Mach number 6 and freestream pressure and temperature 10 kPa and 293 K, respectively. Additionally, high-pressure hydrogen injection near the center of hemispherical blunt nose is simulated using ddtFoam solver \cite{ettner}, which was developed and validated for deflagration to detonation transition in long tubes with blockages. To summarize, the objectives of the current numerical study are: 1) Investigate the effect of sonic hydrogen injection in stagnation zone of hemispherical blunt nose in hypersonic flow-field and compare with equivalent air injection. 2) Study the effects of supersonic injection of hydrogen in the stagnation zone of hemispherical blunt nose in hypersonic flow at different LPM or SPM modes.
\section{Numerical Method}
The numerical simulations are performed by solving unsteady compressible Navier-Strokes equations with $k-\omega$ SST turbulence closure model for an axisymmeytric domain. The ddtFoam \cite{ettner} solver at opensource OpenFOAM platform has been used for all the simulation of air injection as well as hydrogen ignition. The ddtFoam \cite{ettner} utilizes density based approximate Riemann solver with HLLC scheme for solving convective fluxes. The reaction mechanism of hydrogen air combustion is modelled as 9 species 19 reactions O’Conaire reaction mechanism \cite{conaire}. The deflagration and detonation combustion source terms in transport equation of reaction progress variable are evaluated using two look-up tables: reaction products table using mechanism in Cantera \cite{goodwin} and autoignition time delay using Cantera and EDL Toolbox \cite{brown}. The Weller model \cite{weller} has been used for the deflagration combustion modeling and detonation is modelled using auto ignition delay time. It is found that ddtFoam can predict various modes of DDT transition in obstacle laden macro shock-tube. As the ddtFoam solver is developed for modelling accidental fire in large domain of power plants, it utilizes a volume fraction method in each cell to determine the autoignition time-delay, which leads to adequate accurate solutions in coarser grid. On the other hand, to accurately capture the flame propagation, it is required to use 30-50 grid points in the flame thickness region, which may lead to higher computational cost. Hence, it can be said that ddtFoam solver can be utilized for various time dependent extreme combustion or detonation based problems and provide quick results with adequate accuracy for design and analysis purposes. This study further utilizes the ddtFoam solver to study the drag control by using high-pressure hydrogen injection in the stagnation zone. The details of computational domain, boundary conditions and solution procedure are outlines in the following subsections.
\subsection{Computational Domain:}
\begin{figure}[bht]
\centering
\includegraphics[width=0.8\textwidth]{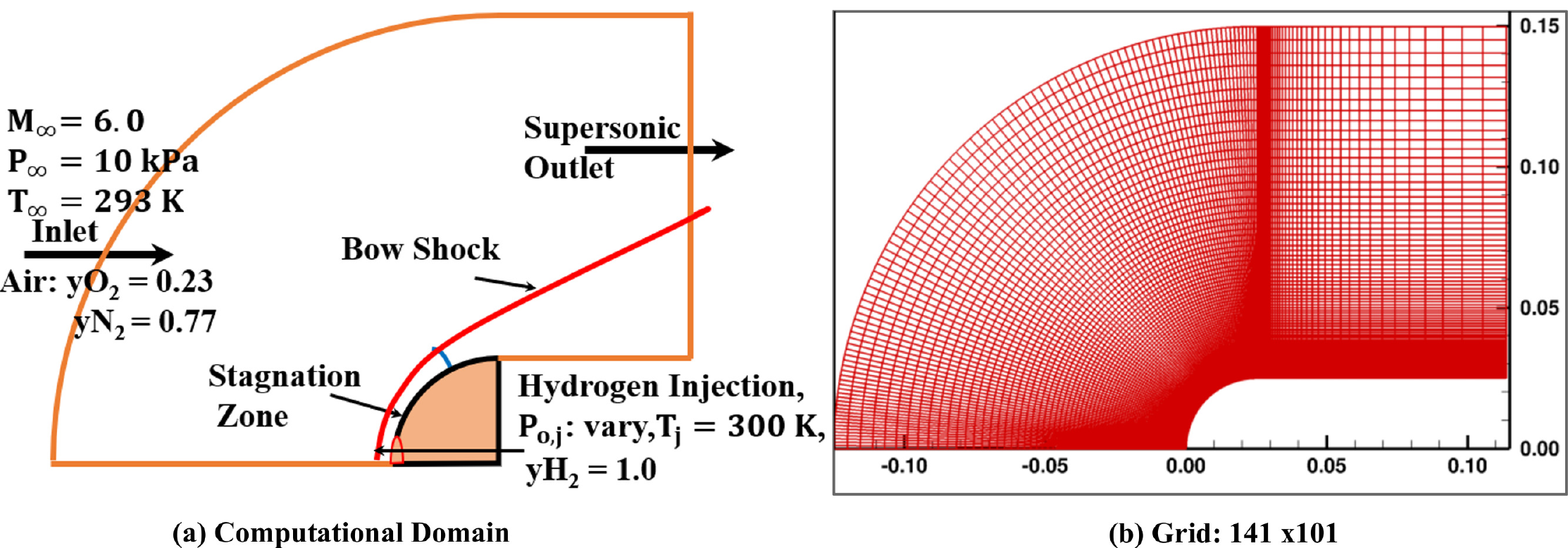}
\caption{\label{Fig01} (a) Computational Domain (b) Axisymmeytric Grid for preliminary study}
\end{figure}
The computational domain consists of a 50 mm diameter hemispherical blunt nose. The counterjet flow is injected through a 4 mm diameter hole at the center of blunt nose. Figure \ref{Fig01}a shows the overall computational domain. The computational domain extends 120 mm after the hemispherical blunt nose to establish the supersonic outlet. The incoming hypersonic flow have been modelled as air with an inlet boundary condition of Mach 6, freestream pressure of 10000 Pa and freestream temperature 293 K. Sonic and supersonic hydrogen or air injected from the center of the hole, which is modelled by imposing the inlet boundary conditions at a 2 mm arc near the center. The temperature of injection jet is maintained at 300 K for all the simulated cases and the stagnation pressure was varied in order to study their counterjet flow effects on drag control. The structured computational grids have been generated as shown in Fig. \ref{Fig01}b. To perform the axisymmetric simulations, the two-dimensional grid have been rotated for $5^{\circ}$ angle with symmetry in the X-Y plane. The wall is modelled as adiabatic wall and the outlet boundary is modelled as a supersonic outlet. The species at the outlet are modelled as a {\it zeroGradient} boundary condition. The numerical simulations for each case has been performed in two steps. Initially, a freestream hypersonic flow is established in front of hemispherical blunt nose by completing the simulation for 1 ms. The hole at the center is considered as adiabatic wall. In the second step, air or hydrogen is injected at sonic or supersonic speed with a temperature of 300 K and varied stagnation pressure. The second simulation is performed for an additional 2 ms physical time, which was sufficient to establish the flow-field in countejet flow.
\subsection{Grid Independence Study:}
\begin{figure}[htb]
\centering
\includegraphics[width=0.5\textwidth]{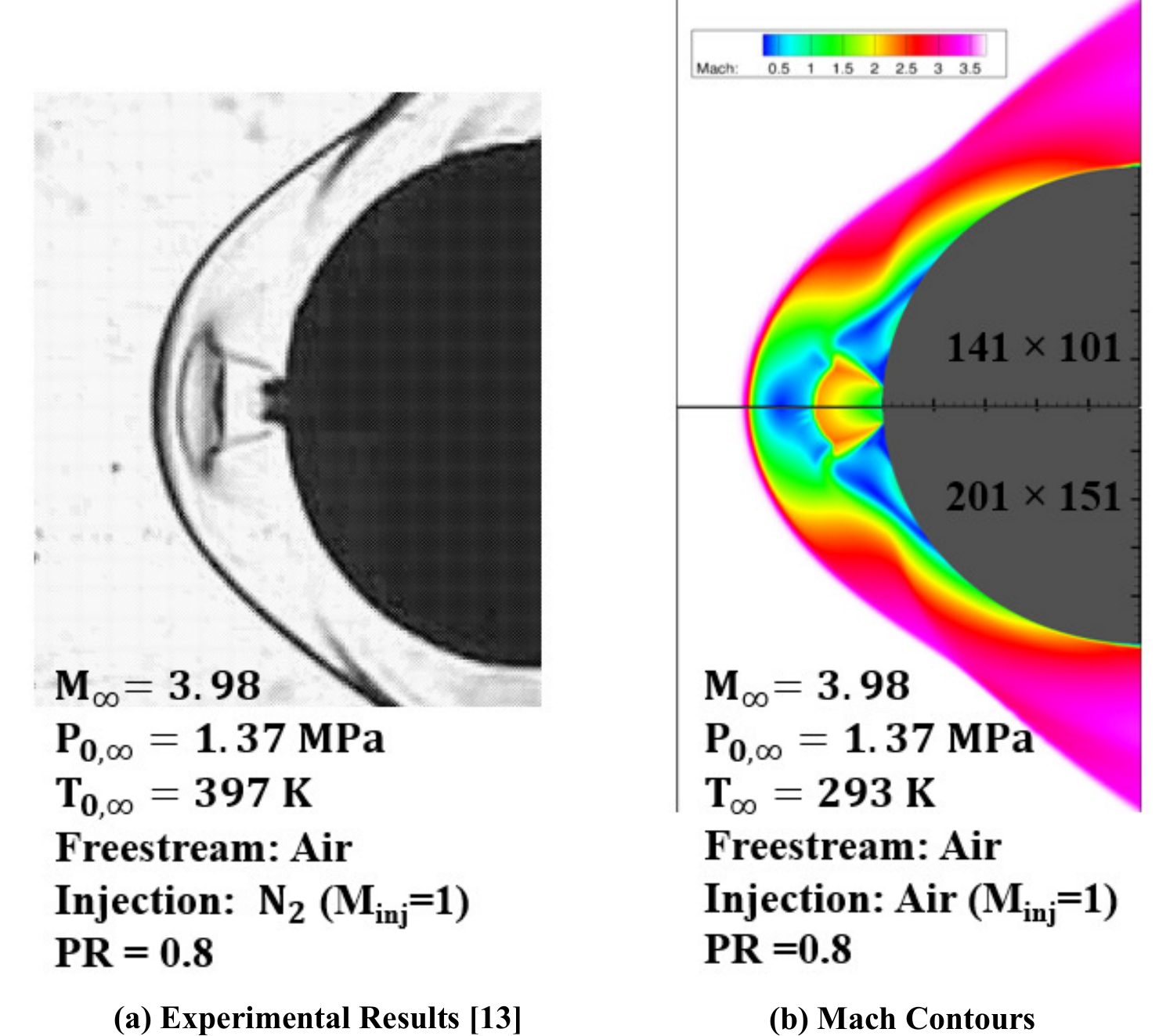}
\caption{\label{Fig02} Air Injection at $PR$= 0.8, (a) Schlieren Visualization \cite{hayashi} (b) Mach contours from ddtFoam for medium and fine grid}
\end{figure}
In order to determine the effectiveness of the ddtFoam solver in counterjet flow application, one of the experimentally studied conditions by Hayashi et al.\cite{hayashi} was simulated for sonic counterjet flow with similar operating condition. The results and operating conditions are shown in Fig.\ref{Fig02}. As the ddtFoam solver was developed for atmospheric or high pressure detonation conditions, the very low pressure and temperature conditions may not be accurately modelled. Hence, in the simulated cases the freestream temperature was changed to 293 K and the sonic injection of air is simulated in place of the $N_2$ injection in experimental case. While the simulated case and experimental case slightly differ in their operating conditions, their flow features can be qualitatively compared. Also, a grid independence study has been performed for the same operating case for three grid sizes and the results are shown in Table 1. Figure\ref{Fig02}a and \ref{Fig02}b shows the qualitative comparison and captured flow-field related to sonic injection for the medium and fine grid configurations. The main flow features such as modified bow shock, Mach disk and the recirculation zones are well captured using the ddtFoam solver for air injection. As this study is mainly focused on drag control, the calculated drag coefficient without flow injection and with sonic injection for three different grid sizes are tabulated in Table 1. The mass flow rate has also been tabulated for all three cases. The drag coefficient for the no injection case shows less than 1\% deviation and with air injection shows less than 2\% deviation when comparing medium and fine grid cases. Hence, medium grid of $141 \times 101$ has been used for all the simulated cases.
%\vspace{-1cm}
\begin{center}
\begin{table}[!htb]
\centering
\caption{\label{jfonts}Grid Independence Study Results.} 
\begin{tabular}{@{}l*{5}{l}}
\br
Grid Size & Drag Coefficient & Drag Coefficient & Mass Flow Rate\\
        & (No Air Injection) & Air Injection (PR = 0.8) & (gms /s)\\
\mr
$71 \times 51$ &0.9736 & 0.6753 & 29.17\\
$141 \times 101$ &0.9788 & 0.6981 & 29.17\\
$201 \times 151$ &0.9797 & 0.6869 & 29.17\\
\br
\end{tabular}
\end{table}
\end{center}
\section{Results and Discussions}
The effectiveness of counterjet flow control depends on several parameters, which have been modelled in many previous studies \cite{reviewp}. The three main parameters for governing performance of counterjet flow drag controls are injection mass flow rate ($\dot{m_j}$), stagnation pressure ratio of injecting jet and freestream flow ($PR$) and momentum ratio ($R_{MA}$) for injecting flow and freestream flow. These parameters can be defined as follows in terms of free stream or injection density, velocity and cross-section area:
\begin{equation}
\dot{m_j} = \rho_{j}V_{j}A_{j}
\end{equation}
where, $\rho_j$, $V_j$ and $A_j$ represent the counterjet exit density, velocity and cross section area, respectively. 
\begin{equation}
 PR = \frac{P_{0,j}}{P_{0,\infty}} 
\end{equation}
where, $P_{0,j}$ and $P_{0,\infty}$ represent the counterjet stagnation pressure and freestream flow stagnation pressure, respectively.
\begin{equation}
 R_{MA} = \frac{\rho_{j}V_{j}^2 A_{j}}{\rho_{\infty}V_{\infty}^2A_{\infty}}
\end{equation}
additionally, here $\rho_{\infty}$, $V_{\infty}$ and $A_{\infty}$ represent the freestream density, freestream velocity and cross section area of blunt nose, respectively. \\
The total drag in each simulations have been calculated by using the following equation:
\begin{equation}
 C_{D,Total} = C_{D,A} + \dot{m_j}V_j + (P_j-P_{\infty})A_j
\end{equation}
here, $C_{D,Total}$ is total drag force experienced by blunt nose, $C_{D,A}$ is drag force experienced by blunt nose (pressure + viscous), and the next two terms represent drag due to thrust generated by counterjet flow.
In this study, the freestream conditions for air have been kept fixed as hypersonic Mach number $M_{\infty} = 6$, freestream pressure $P_{\infty} = 10000$ Pa and temperature $T_{\infty} = 293$ K. The cross-sectional area of hemispherical blunt nose $A_{\infty}$ and cross-sectional area of counterjet exit $A_{j}$ are also fixed. The counterjet exit temperature, $T_{j} = 300$ K has also been kept fixed. It should be noted that the drag reduction can be achieved by counterjet flow, when the momentum ratio is significantly less than 1 in order to generate the least thrust from the counterjet. If the momentum ratio is higher than 1, it will produce high thrust and work as retro-propulsion. The stagnation pressure ratio $PR$ signifies the strength of a counterjet flow, it also governs the mass flow rate of a counterjet. In this study, when the comparison of the effectiveness of a counterjet flow between air and hydrogen was made, $PR$ and $R_{MA}$ were kept constant, which lead to a smaller mass flow rate for hydrogen injection than air and higher exit velocity for hydrogen than air. The results are discussed for sonic ($M_j = 1$) injection of hydrogen and air, and supersonic injection ($M_j = 2.94$)of hydrogen and air. The sonic counterjet injection cases for air and hydrogen injections have been compared for a single $PR = 0.12$, which shows the flow-features as short penetration mode. Further more supersonic counterjet scenarios have been compared for air and hydrogen injection for different $PR$ values in order to simulate short and long penetration modes.

\subsection{Sonic Air and hydrogen Injection:}
\begin{figure}[htb]
\centering
\includegraphics[width=0.8\textwidth]{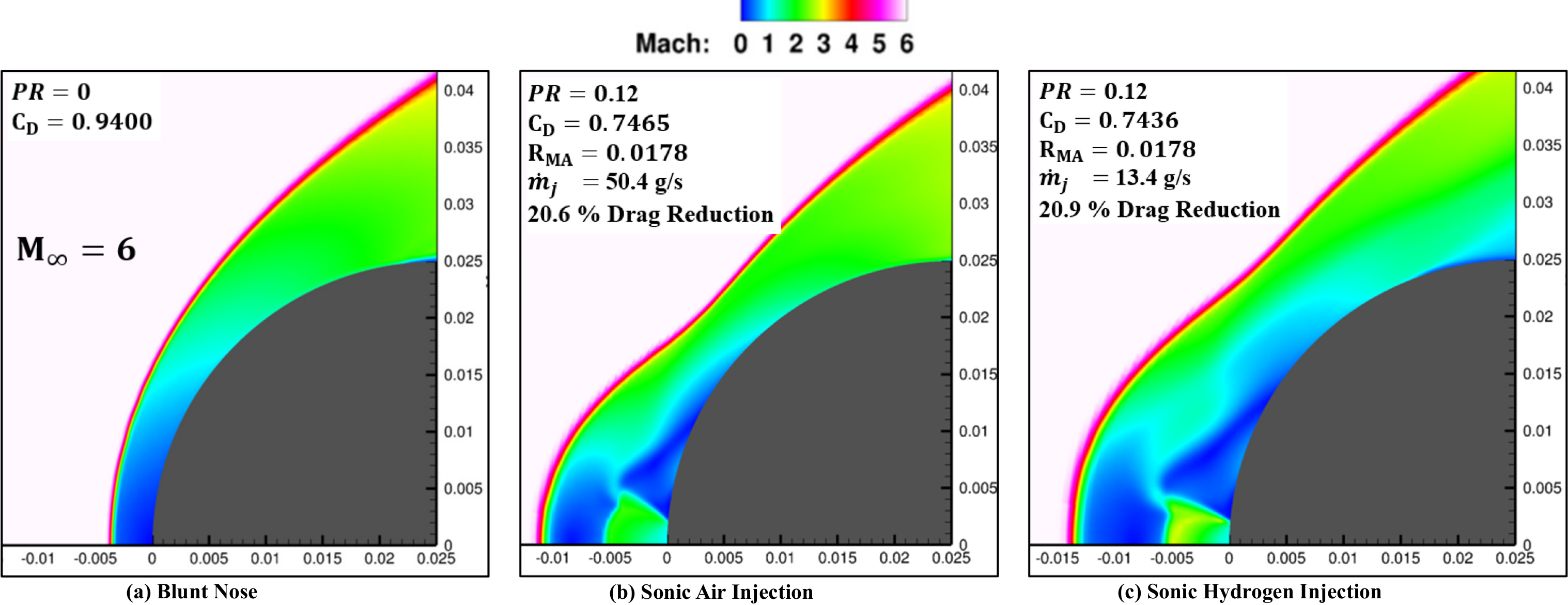}
\caption{\label{Fig03} Mach contours for (a) Blunt Nose (b) Sonic Air Injection (c) Sonic Hydrogen Injection}
\end{figure}

Figure 3 shows Mach contours for a blunt nose without any counterjet injection in the freestream air with a Mach number 6. The computed drag coefficient for the hemispherical blunt nose is 0.94. Figure \ref{Fig03}b and \ref{Fig03}c show the flow features for air and hydrogen counterjet flows exiting from the sonic nozzle at $PR$= 0.12. For both cases, the momentum ratio remains constant as 0.0178. For this stagnation pressure ratio, the air and hydrogen counterjet flows show similar flow features in their Mach contours. However, the bow shock at the stagnation line moves slightly away from body for hydrogen injection in comparison to air injection. However, there is no significant effect on drag reduction, as both the gas injection scenarios lead to approximately a 21 \% drag reduction. However, the required mass flow rate for hydrogen is almost 3.76 times less than the required mass flow rate of air to achieve the same performance.
\begin{figure}[htb]
\centering
\includegraphics[width=0.9\textwidth]{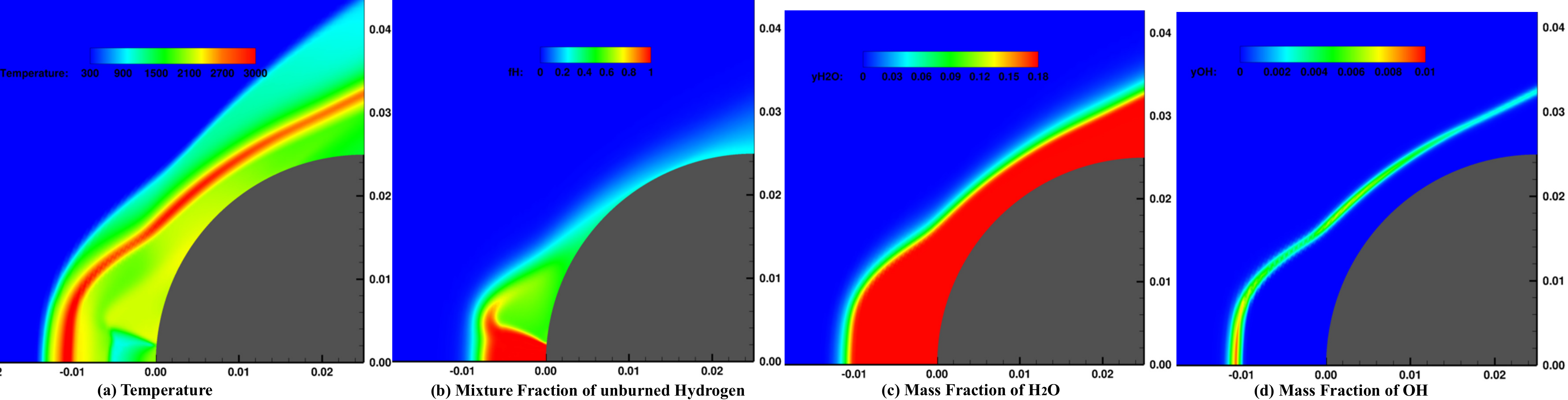}
\caption{\label{Fig04} (a) Temperature, (b) Hydrogen Mixture Fraction (c)$H_{2}O$ Mass Fraction, (d) OH Mass Fraction for sonic hydrogen injection case}
\end{figure}
Fig. \ref{Fig04} shows the contour plots for temperature, hydrogen mixture fraction, $H_{2}O$ mass fraction and $OH$ mass fraction for hydrogen injection by the sonic nozzle at $PR$= 0.12. The mixture fraction was modelled as unburned fuel mass fraction if the reaction is not present \cite{ettner}. The temperature plot shows a high temperature zone away from the blunt body, which can be clearly seen as a standing flame in the $OH$ mass fraction plot. The $H_{2}O$ mass fraction shows that the reaction product can completely submerge the blunt body. The main finding is that the counterjet of a lighter or heavier gas may provide similar levels of drag control in SPM modes for the same momentum ratio and PR, but the lighter gas requires a lower mass flow rate resulting in reduced storage requirements.
\subsection{Supersonic Air and hydrogen Injection:}
\begin{figure}[htb]
\centering
\includegraphics[width=0.9\textwidth]{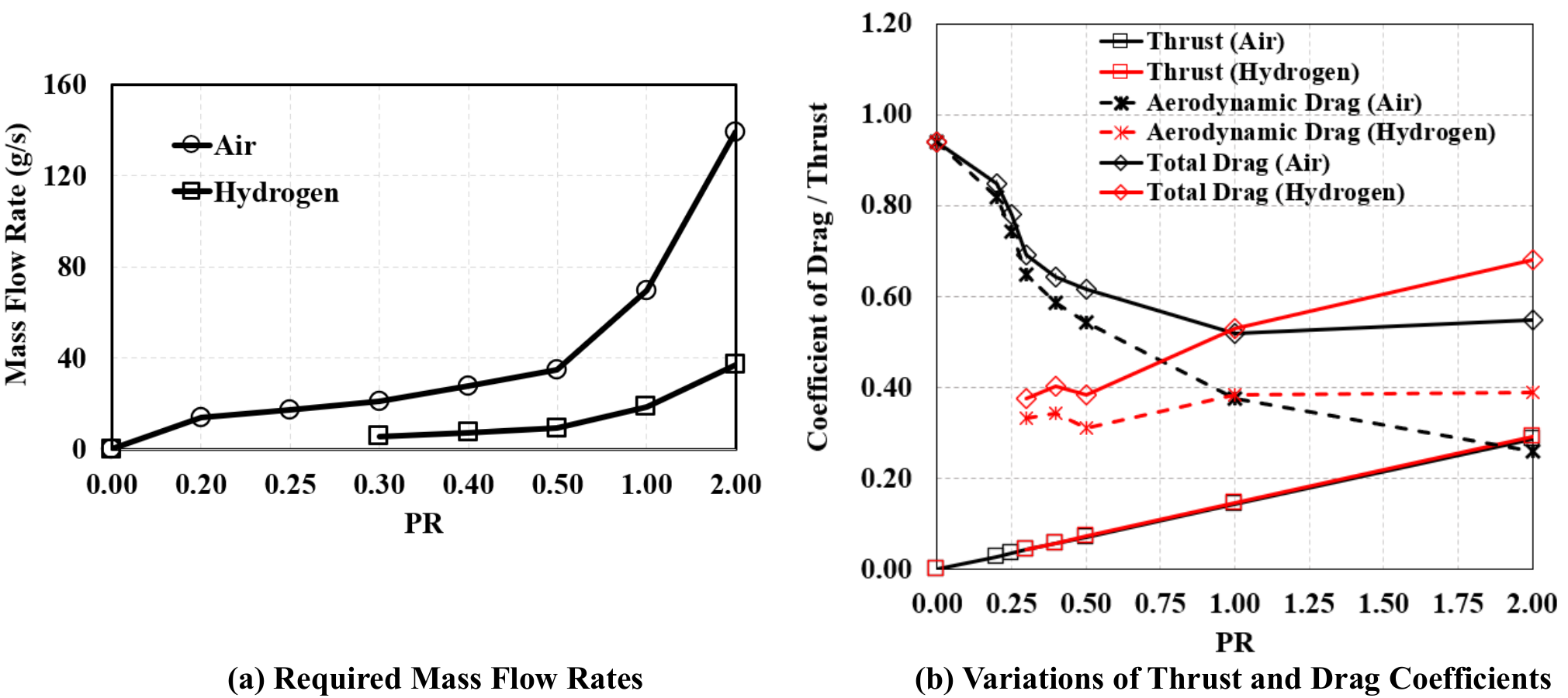}
\caption{\label{Fig05} (a) Mass Flow Rates (b) Thrust and Drag Coefficient variations with change in $PR$ for supersonic ($M_j = 2.94$)air or hydrogen injection}
\end{figure}
\vspace{-1cm}
\begin{center}
\begin{table}[h]
\centering
\caption{\label{table2} Observed behaviour of Supersonic Counterjet flow.} 
\begin{tabular}{@{}l*{6}{l}}
\br
PR & 0.20 & 0.30 & 0.40 & 0.50 & 1.0 & 2.0\\
\mr
Air & LPM & LPM & LPM & LPM & SPM & SPM \\
Hydrogen & Unstable & Oscillatory & Oscillatory & Oscillatory & Intermittent & Intermittent \\
         & Solution & LPM & LPM & LPM & SPM-LPM & SPM-LPM \\
\br
\end{tabular}
\end{table}
\end{center}
The sonic counterjets generally behave in a short penetration mode. In order to achieve a long penetration mode with a wide range of operating pressures, the use supersonic nozzle is required. The supersonic injection of air or hydrogen (at $M_j = 2.94$) have been modelled as inlet conditions at the center hole location of a blunt nose with same diameter as 4 mm. The simulations were performed by changing the stagnation pressure ratio from $PR$= 0.25 to 2.0 as shown in Fig. \ref{Fig05}. The momentum ratio $R_{MA}$ for each $PR$ have been constant for air and hydrogen injection, which results in a 3.76 times lower mass flow rate for hydrogen injection in comparison to air injection (Fig. \ref{Fig05}a). The hydrogen injection for $PR$ less than 0.3 were not stable during the simulation, hence were not included in the results. The time-averaged aerodynamic drag (including pressure and viscous drag on the blunt nose), drag due to thrust by counterjet flow and total drag in each case of air and hydrogen injection are plotted in Fig. \ref{Fig05}b. Table \ref{table2} summarises the behaviour of counterjet for both air and hydrogen injection. It can be seen that the drag due to thrust from the counterjet increases linearly as $PR$ increases for both hydrogen and air, because the $R_{MA}$ remains constant for each $PR$. In the case of air injection, the aerodynamic drag decreases with an increase in PR and the counterjet changes its behaviour from LPM to SPM. However, at higher $PR$, the thrust also increases, which leads to an over-all increase in the total drag at higher $PR$. For hydrogen injection, $PR$ = 0.3, 0.4 and 0.5 exhibit a long penetration mode and a drag reduction up to 60 \%. The flow-field is oscillatory in nature. At higher $PR$ = 1 and 2 the aerodynamic drag increases as flow-field changes to SPM for a longer time and intermittently change to LPM mode because of the high pressure burning. It can be concluded that at low $PR$, operating at a smaller mass flow rate, hydrogen injection provides higher drag reduction by pushing the shock wave far away from blunt body. A future heat transfer study will provide a clear idea of the overall effectiveness of hydrogen injection in LPM mode for drag as well as aerodynamic heating control.

\section{Conclusions}
The effectiveness of a hybrid drag control method for a sonic and supersonic hydrogen injection in the stagnation zone of a blunt nose at hypersonic Mach number 6 have been numerically investigated. The sonic hydrogen injection lead to a SPM flow-field, similar to air injection in front of the blunt nose and provide same drag reduction performance at 26 \% lower mass flow rate. The supersonic hydrogen injection behaves differently to the supersonic air injection, leading to a higher drag reduction up to 60 \% at lower $PR$ (= 0.3 to 0.5) and mass-flow rate as low as 5.6 - 10 g/s. At higher $PR$ values, air injection performs better than hydrogen injection, but may require very high mass flow rates up to 140 g/s. Future work will focus on a heat transfer study to analyse the effects of hydrogen injection on drag as well as aerodynamic heating at LPM modes.

\smallskip
\end{document}